\documentclass[fleqn,twocolumn,10pt]{wlscirep}
\usepackage{color}
\usepackage{graphicx}
\usepackage{dcolumn}
\usepackage{array}
\usepackage{float}
\usepackage{longtable}
\usepackage{mathrsfs}
\usepackage{txfonts}
\usepackage{wasysym}
\usepackage[T1]{fontenc}
\usepackage{amssymb}

\def\la{\lambda}

\def\t{{\rm t}}

\def\la{\langle}
\def\ra{\rangle}

\def\tr{{\rm Tr}}

\newcommand{\beq}{\begin{equation}}
\newcommand{\eeq}{\end{equation}}
\newcommand{\beqa}{\begin{eqnarray}}
\newcommand{\eeqa}{\end{eqnarray}}

\newcommand{\ket}[1] {\vert #1 \rangle}
\newcommand{\bra}[1] {\langle #1 |}

\newcommand{\mix}{\mathrm{mix}}
\newcommand{\Th}{\mathrm{th}}

\begin{document}

\title{Quantum work statistics, Loschmidt echo and information scrambling}

\author[1,2,*]{A. Chenu}
\affil[1]{Massachusetts Institute of Technology,
77 Massachusetts Avenue, Cambridge, MA 02139, USA}
\affil[2]{Theoretical Division, Los Alamos National Laboratory, Los Alamos, New Mexico 87545, USA}

\author[3]{I. L. Egusquiza}
\affil[3]{Department of Theoretical Physics and History of Science,\\ University of the Basque Country UPV/EHU, Apartado 644, 48080 Bilbao, Spain}
\author[4]{J. Molina-Vilaplana}
\affil[4]{Technical University of Cartagena, UPCT, 30202, Cartagena, Spain}
\author[5]{A. del Campo}
\affil[5]{Department of Physics, University of Massachusetts, Boston, MA 02125, USA}

\affil[*]{achenu@mit.edu}

\begin{abstract}
A universal relation is established between  the quantum work probability distribution of an isolated driven quantum system and  the Loschmidt echo dynamics of a two-mode squeezed  state.  When the initial density matrix is canonical, the Loschmidt echo of the purified double thermofield state provides a direct measure of information scrambling and can be related to the analytic continuation of the partition function. Information scrambling  is then described by the quantum work statistics associated with the  time-reversal operation on a single copy, associated with  the sudden negation of the system Hamiltonian.

\end{abstract}

\flushbottom
\maketitle

\section*{Introduction}
Quantum thermodynamics provides a framework to unify quantum theory, statistical mechanics, information theory and thermodynamics \cite{Vinjanampathy2016a}.
A central object in this field is the notion of work associated with the dynamics of an isolated quantum system.
At the quantum level, work becomes a stochastic variable, described by a probability distribution \cite{Talkner2007a}.
The analysis of the associated work statistics has guided seminal developments in stochastic thermodynamics and  nonequilibrium statistical mechanics. Prominent examples  include the  Jarzynski equality \cite{Jarzynski1997a,Tasaki2000a,Kurchan2000a} and fluctuation theorems \cite{Crooks1999a,Seifert2012a}, with both classical and quantum counterparts \cite{Campisi2011a}.
Quantum thermodynamics 
is strongly tied to quantum dynamics and the notion of reversibility. The sensitivity of  a quantum system to external perturbations is often characterized via a Loschmidt echo that measures the extent to which quantum evolution can be reversed upon an imperfect time-reversal operation \cite{Gorin2006a,Jacquod2009a, Goussev2012a}.
In this context, the generating function of the work probability distribution associated with the driving of a pure energy eigenstate via a quantum quench is known to be identical to the Loschmidt echo of such eigenstate \cite{Silva2008a}.  This observation has greatly facilitated the understanding of quantum work fluctuations in finite-time, nonequilibrium thermodynamics of many-body systems \cite{Silva2008a,Abeling2016a,Garcia-March2016a,Campbell2016a}.
More recently, it has been suggested that the work probability distribution is related to the out-of-time order correlators (OTOC) \cite{Campisi2017a,YungerHalpern2017a}.  First discussed  in condensed matter physics \cite{Larkin1969a} and the study of   irreversible processes \cite{Talkner1979a}, OTOC are currently under exhaustive investigation  to diagnose quantum chaos and scrambling of information in black hole physics \cite{Maldacena2016a}. 

This Report establishes a fundamental connection between work statistics, Loschmidt echo, and information scrambling. 
This is done by first showing that  the work statistics associated with an arbitrary driving protocol of a generic quantum state is equivalent to the  Loschmidt echo dynamics of a  two-mode squeezed state in an enlarged Hilbert space. 
For an initial thermal state,  the two-mode squeezed state becomes a thermofield double state, as that used to describe eternal black holes \cite{Maldacena2003a}.  The work probability distribution resulting from a perfect time-reversal operation on a given system is determined  by  the analytic continuation of the partition function, and then links to information scrambling. More generally, we show that work statistics dictates information scrambling resulting from an arbitrary  Loschmidt echo.

\section*{Results}
\subsection*{Universal relation between quantum work statistics and Loschmidt echo dynamics}Let us consider an isolated quantum system in a Hilbert space $\mathcal{H}$ and described by the time-dependent Hermitian Hamiltonian $\hat{H}_s=\sum_nE_n^s|n_s\ra\la n_s|$, with  instantaneous eigenstates $ |n_s\ra$  and eigenenergies $E_n^s$. 
We consider the  evolution from time $s=0$  to time $s=\tau$ of an arbitrary initial state with density matrix  $\hat{\rho}$, 
dictated by the  evolution operator 
\beqa
\label{Utau}
\hat{U}(\tau)={\cal T}\exp\left[-i\int_0^\tau\hat{H}_sds\right], 
\eeqa
where ${\cal T}$ is the time-ordering operator. We include here also the case of sudden quenches, with $\tau\to0^+$ such that $\hat{U}(\tau)\to\mathbf{1}$, but in which the final Hamiltonian, for which we retain the notation $\hat{H}_\tau$, is different from the initial hamiltonian $\hat{H}_0$.

Characterizing the work done during this driving protocol requires  two projective energy measurements: one at the initial time $s=0$ and another at $s=\tau$ \cite{Kurchan2000a,Talkner2007a}. The results of both measurements, respectively $E_n^0$ and $E_m^\tau$, give the work done as $W=E_m^\tau -E_n^0$. This two-energy measurement scheme
prevents work from being defined as an observable in the quantum world \cite{Talkner2007a}. Even so, it  can be understood in terms of a generalized measurement scheme \cite{Mazzola2013a,Roncaglia2014a}. The  corresponding work probability distribution can be written as ~\cite{Kurchan2000a, Tasaki2000a}
\begin{equation}
p(W):=\sum_{n,m} p^0_{n}\;  p^\tau_{m \vert n} \delta\left[W-\left(E_m^\tau -E_n^0\right)\right],
\label{defwork}
\end{equation}
where  $p^0_n = \langle n_0|\hat{\rho}|n_0\rangle$ is the probability that the initial state  is found in the $n$-th eigenstate  of the initial Hamiltonian, and $p^\tau_{m \vert n}\geq0$ is the transition probability from this initial eigenstate to the $m$-th eigenstate of the final Hamiltonian $|m_\tau\ra$, 
i.e.,
\beqa \label{eq:pmn}
p^\tau_{m \vert n}=|\la m_\tau|\hat{U}(\tau)|n_0\ra|^2.
\eeqa
The integral representation of the delta function in terms of an auxiliary variable $t$, $\delta(W-E)=\frac{1}{2\pi}\int_{-\infty}^\infty dt \, e^{-it(W-E)}$, allows us to write the work distribution as the Fourier transform
\beqa
\label{pWIF}
p(W)=\frac{1}{2\pi}\int_{-\infty}^\infty dt \, \chi(t,\tau)e^{-itW}
\eeqa
of the characteristic function
\beqa \label{eq:chi_def}
\chi(t,\tau)&=&\sum_{n,m} p^0_n\;  p^\tau_{m \vert n}e^{it\left(E_m^\tau-E_n^0\right)}.
\eeqa
 We show below that a universal relation exists between the work probability distribution $p(W)$ in an arbitrary unitary protocol and the dynamics of a  Loschmidt echo, for any initial state, including mixed states, e.g. at finite temperature.

Notice that even if the initial state  is pure,  the first projective energy measurement in the initial eigenbasis 
generally leads to a (post-measurement) mixed state
\begin{equation} \label{eq:rhomix}
\hat{\rho}_\mix= \sum_n \hat{P}_n^0 \: \hat{\rho} \: \hat{P}_n^0 = \sum_{n} p^0_n \ket{n_0}\bra{n_0},
\end{equation} 
where  $\hat{P}_n^0$ denotes  the projector on the $n$-th energy eigenstate of the initial Hamiltonian at $s=0$.
Using the explicit definition of the transition probability, Eq. (\ref{eq:pmn}), the characteristic function (\ref{eq:chi_def})  
can then be written as
\beqa \label{characteristicfunction}
\chi(t,\tau)&=&\tr \left( \hat{U}^\dag(\tau) e^{it\hat{H}_\tau} \hat{U}(\tau) \, e^{-it\hat{H}_0}\hat{\rho}_\mix \right).
\eeqa
This form allows us to identify the auxiliary variable $t$ as a second time of evolution, different from $s$, as was first proposed in Ref.  \cite{Silva2008a}. Notice that the mixed state $\hat{\rho}_{\mathrm{mix}}$ is stationary with respect to $\hat{H}_0$, which entails the property $\chi^*(t,\tau)=\chi(-t,\tau)$.

Whenever the initial state is an energy eigenstate of $\hat{H}_0$, $\ket{j_0}$, the initial density matrix simplifies to $\hat{\rho}=\hat{\rho}_\mix = \ket{j_0}\bra{j_0}$ and $p^0_n =\la n_0|\hat{\rho}|n_0\ra= \delta_{nj}$. In the sudden quench limit  $\tau\rightarrow 0^+$, the transition probability $p^\tau_{m \vert n}$ further reduces to $|\la m_\tau|n_0\ra|^2$, and the characteristic function becomes 
\beqa
\label{chi:Losch}
 \chi(t,0^+)=\la j_0|e^{+it\hat{H}_\tau}e^{-it\hat{H}_0}|j_0\ra\,,
\eeqa
recognizable 
as a Loschmidt amplitude $A(t)$,  
given by the survival amplitude of the initial eigenstate $|j_0\ra$ first propagated forward in time with $\hat{H}_0$ and subsequently backward in time with a dynamics generated by $\hat{H}_\tau$ \cite{Silva2008a}.  
Thus we are presented with the Loschmidt amplitude for  a quantum quench in the $t$ evolution, with a driving protocol of the form 
\beqa \label{eq:tau0}
\hat{H}(t)=\hat{H}_0\, \Theta(-t)-\hat{H}_\tau\, \Theta(t),
\eeqa
where $\Theta(t)$ is the Heaviside function. 
The survival probability  
$\mathcal{L}(t)=|A(t)|^2$
is known as a Loschmidt echo, 
 and can be further related to the local density of states  \cite{Wisniacki2002a}.  While its exponential decay in time has often been used to characterize chaotic systems  \cite{Gorin2006a}, it can occur in simpler, integrable systems \cite{Dubertrand2014a}.

 In fact, the relation between the generating function and  the Loschmidt echo amplitude is  universal, since it holds for any generic state -pure or not- under any driving protocol. To show this,  we proceed in two steps.
First, we consider general  initial states $\hat{\rho}$ that give rise to the post-measurement  state $\hat{\rho}_{\rm mix}$. This state $\hat{\rho}_{\rm mix}$ is then purified by  embedding it in the extended Hilbert space $\mathcal{H}_L\otimes\mathcal{H}_R$, with $\mathcal{H}_L = \mathcal{H}_R =\mathcal{H}$, e.g. defining the double-copy state  
\beqa
\label{pstate}
|\Psi_0\ra=\sum_n\sqrt{p^0_n}\,
\,|n_0\ra_L\otimes|n_0\ra_R. 
\eeqa
%
Secondly, we introduce an effective single-copy Hamiltonian $\hat{H}_\tau^{\rm eff}$, acting on $\mathcal{H}$,
\begin{equation}
  \label{eq:heffective}
  \hat{H}_\tau^{\rm eff}=\hat{U}^\dagger(\tau)\hat{H}_\tau\hat{U}(\tau)\,.
\end{equation}
 We note that the dynamics associated with $ \hat{H}_\tau^{\rm eff}$ with an initial pure state has been recently discussed in \cite{Garcia-Mata2017a,Garcia-Mata2018a}.

As a result, the characteristic function resulting from the evolution $\hat{U}(\tau)$ in  Eq. (\ref{Utau}) is equal to the Loschmidt echo amplitude
 $A(t)=\la\Psi_0|\Psi_t \ra$ of the purified state (\ref{pstate}), when one of the copies evolves under the sudden quench 
\begin{equation} \label{genquench}
\hat{H}(t)=\hat{H}_0\, \Theta(-t)-\hat{H}_\tau^{\rm eff}\, \Theta(t).
\end{equation}
Denoting the composition of unitary evolution operators  by the echo matrix $ \hat{M}(t,\tau)\equiv  e^{+it\hat{H}_\tau^{\rm eff}}e^{-it\hat{H}_0}$, 
the characteristic function (\ref{characteristicfunction})  becomes 
\beqa
\label{chigen}
\chi(t,\tau)=\la\Psi_0| \hat{M}(t,\tau) \otimes \mathbf{1}_R |\Psi_0\ra,
\eeqa
where we have emphasized that the evolution acts exclusively on one of the copies, e.g. here, the left one.
In particular, (\ref{chigen}) corresponds to the echo dynamics of a  purified initial state that evolves first forward in time under the initial Hamiltonian $\hat{H}_0$ and  then backward  under the effective final Hamiltonian  $\hat{H}_\tau^{\rm eff}$, while the second copy is left unchanged.  Explicitly,
 \begin{equation} \label{chi=A}
\chi(t,\tau)  = A(t).
\end{equation}
This result shows that the equivalence between the Loschmidt amplitude and the work statistics holds in a universal setting, and extends the correspondance that was first demonstrated for a system  prepared in an energy  eigenstate of the initial Hamiltonian $\hat{H}_0$ undergoing a sudden quench  in  \cite{Silva2008a}. 


As direct consequence of this equivalence, the short-time decay of the Loschmidt echo is determined by the variance of the quantum work statistics. In particular,  the Loschmidt echo is an even function since ${\cal L}(t)=\left|\chi(t,\tau)\right|^2=\chi(t,\tau)\chi(-t,\tau)$. Under the assumption of analyticity of the generating function of the cumulants $C_n$ of the work probability density, we have $\ln\left[\chi(t,\tau)\right]=\sum_{n=1}^\infty(i t )^n C_n / {n!}$, which entails
  \begin{eqnarray}
    \label{eq:loschexpansion}
   \ln {\cal L}(t)&=&\ln\left|\langle \Psi_0| \hat{M}(t,\tau)\otimes \mathbf{1}_R|\Psi_0\rangle\right|^2\nonumber\\
                  &=&2\sum_{n=1}^\infty\frac{(-1)^n t^{2n}}{(2n)!} C_{2n}\nonumber\\
                  &=& -\Delta W^2t^2+\mathcal{O}(t^4)    \,,
  \end{eqnarray}
in terms of the variance of the work fluctuations $C_2=\la W^2\ra-\la W\ra^2\equiv\Delta W^2$ associated with the quench (\ref{genquench}) on the system.
This universal short-time asymptotics is reminiscent of the well-known short-time quadratic decay under unitary evolution and that gives rise to the quantum Zeno effect 
\cite{Chiu1977a},  with a Gaussian width  now identified as the work fluctuations. 


\subsection*{Quantum work statistics of time reversal and information scrambling}
As the connection in Eq. (\ref{chi=A}) is valid  for any generic state, it applies in particular to mixed states at finite temperature. 
In what follows, we shall consider the initial state to be the canonical thermal state  in  $\mathcal{H}$,  with density matrix
\beqa
\label{rhothermal}
\hat{\rho}=\hat{\rho}_\Th=\frac{e^{-\beta\hat{H}_0}}{\tr \left(e^{-\beta\hat{H}_0 } \right)}, 
\eeqa
where $\beta = (k_B T)^{-1}$ and $T$ is the temperature. In this case, $\hat{\rho}=\hat{\rho}_{\mathrm{mix}}$, and   the purification of both  is the  so-called thermofield double state \cite{Semenoff1983a}, 
 \beqa
\label{TDS}
|\Psi_0\ra=\frac{1}{\sqrt{Z(\beta)}}\sum_ne^{-\frac{\beta}{2} \hat{H}_0\,\otimes\, \mathbf{1}_R }\,\,|n_0\ra_L\otimes|n_0\ra_R,
\eeqa
 where $\hat{H}_0$ acts on only one of the copies and the normalization factor $Z(\beta)=\tr_{\mathcal{H}} [e^{-\beta\hat{H}_0 } ] = \sum_n e^{-\beta E_n^{0}}$  is the partition function.
 
 We shall focus on the work statistics associated with the implementation of the  time-reversal operation in the original system in $\mathcal{H}$. 
 When the system Hamiltonian is time-reversal, the implementation of the time-reversal operation is  equivalent to the sudden negation of the Hamiltonian according to the quench
\beqa
\label{PTR}
\hat{H}(t)&=&\hat{H}_0\, \Theta(-t)-\hat{H}_0\, \Theta(t). 
\eeqa
In the laboratory, such negation of the Hamiltonian can be implemented  making use of a control ancilla $C$, when the dynamics of the ancilla-system in the Hilbert space $\mathbb{C}^2\otimes\mathcal{H}$ is generated by the Hamiltonian $\sigma_C^z\otimes \hat{H}_0$, where $\sigma_C^z$ is the Pauli matrix acting on the ancilla \cite{Swingle2016a,Garcia-Alvarez2017a}.

 The  generating  function of the corresponding work statistics can be explicitly computed to be
 \beqa
 \label{chiTDS}
 \chi(t,0^+)
 &=&\frac{1}{Z(\beta)}\sum_ne^{-(\beta+2it) E_n^{0}}\nonumber\\
&=&\frac{Z\left(\beta + i2t  \right)}{Z\left(\beta \right)},
 \eeqa
 where we have  used the  analytically-continued partition function. 
 The equivalence established above gives the  Loschmidt echo  $\mathcal{L}(t)=|\chi(t,0^+)|^2$  as the fidelity between the initial thermofield double state $|\Psi_0\ra$ and its time-evolution, also known as the survival probability:
 \beqa
 \label{letds}
\mathcal{L}(t)=|\la\Psi_0|\Psi_t\ra|^2&=&\left|\frac{Z\left(\beta + i2t  \right)}{Z\left(\beta \right)}\right|^2. 
\eeqa
This observation leads to a direct connection with information scrambling in black hole physics \cite{Hayden2007a,Sekino2008a,Barbon2003a,Barbon2004a}, where the  thermofield double state represents an entangled state of two conformal field theories that is dual to an eternal black hole via the AdS/CFT correspondence \cite{Maldacena2003a}.
Specifically, an eternal black hole, which amounts to a non-traversable wormhole between two asymptotic regions of spacetime, has been conjectured to be equivalent to a pair of entangled black holes in a disconnected space with common time \cite{Maldacena2013a}.
In this interpretation, the thermofield double state (\ref{TDS}) is not invariant under time-evolution.
Further, it has been argued that no system scrambles information faster than a black hole \cite{Sekino2008a,Maldacena2016a}, 
prompting the analysis of various tools to diagnose chaos, including the fidelity decay of a thermofield double state, Eq. (\ref{letds}) \cite{Papadodimas2015a,Cotler2017b,Dyer2017a,delCampo2017a}.
Therefore, the work statistics associated with the implementation of the time-reversal operation in a system dictates the dynamics of information scrambling as described by the survival probability of the thermofield double state.

Alternatively,  information scrambling can also be related to the work probability distribution associated with a quench in the enlarged Hilbert space $\mathcal{H}\otimes\mathcal{H}$ described by the Hamiltonian 
\beqa
\label{TRinHH}
\hat{H}(t)\otimes \mathbf{1}_R+\mathbf{1}_L\otimes \hat{H}_0,
\eeqa
with $\hat{H}(t)$ given in Eq. (\ref{PTR}), as
\beqa
\frac{Z\left(\beta + i2t  \right)}{Z\left(\beta \right)} = \bra{\Psi_0} e^{ i t \left( - \hat{H}_0\otimes \mathbf{1}_R +\mathbf{1}_L\otimes \hat{H}_0\right)} e^{ -i t \left(\hat{H}_0\otimes \mathbf{1}_R +\mathbf{1}_L\otimes \hat{H}_0\right)} \ket{\Psi_0}.
\eeqa
This corresponds to a local time-reversal operation acting on the left copy only, leaving the right one untouched. In fact any independent unquenched evolution for the right copy would give the same result. 
 

The work probability distribution is the same in both interpretations and follows from  (\ref{pWIF})    
\beqa
p(W)&=&\frac{1}{ Z(\beta)}\tr\, \left[e^{-\beta\hat{H}_0 }\delta\left(W+2\hat{H}_0\right)\right] \nonumber\\
&=&\left\la \delta\left(W+2\hat{H}_0\right) \right\ra_\beta,
\eeqa
where $\left\la\hat{A} \right\ra_\beta = \tr_\mathcal{H}\left(\hat{A} \, \hat{\rho}_\Th\right)$ denotes the thermal average of $\hat{A}$ in the canonical ensemble. Therefore, the work probability distribution for a time-reversal operation  is given by the thermal average of the density of states operator, $\rho(E)= \delta\left(\hat{H}_0-E\right)$,  identifying $E$ with $-W/2$, i.e.,
\beqa
p(W)=\frac{1}{2}\left\la\rho(E)\right\ra_\beta\big\vert_{E=-W/2}.
\eeqa
As a result, it follows that the mean work done to reverse the dynamics of one of the entangled copies in the thermofield double state $|\Psi_0\ra\in\mathcal{H}\otimes\mathcal{H}$ can be obtained from the mean thermal energy of the canonical ensemble, in $\mathcal{H}$, 
\beqa
\la W\ra=\int dW Wp(W)= - 2 \la \hat{H}_0\ra_\beta.
\eeqa

Conversely, a number of protocols are available to measure $p(W)$ \cite{Dorner2013a,Mazzola2013a,Roncaglia2014a} and have been successfully implemented in the laboratory \cite{Batalhao2014a,An2015a,Cerisola2017a}. 
The survival amplitude of the thermofield double state, $A(t)=Z\left(\beta + i2t  \right)/Z\left(\beta \right)$, can thus be accessed from a measurement of the work statistics $p(W)$ associated with a time-reversal operation, that via inverse Fourier transform yields 
$A(t)=\chi(t,0^+)=\int dWp(W)e^{+itW}$. The Loschmidt echo simply follows as 
\beqa \label{LEequivalence}
\mathcal{L}(t)=|A(t)|^2=|\chi(t,0^+)|^2.
\eeqa
These relations result from the implementation of a perfect time-reversal operation described by the quench (\ref{PTR}) or (\ref{TRinHH}).
More generally, one is led to consider an arbitrary quench dynamics on one of the copies, that can accommodate for imperfect time-reversal operations,  driven by a quench from $\hat{H}_0$  to  $\hat{H}_\tau$. The associated Loschmidt echo reads
\beqa
\mathcal{L}(t)&=&|\chi(t,\tau)|^2,\nonumber\\
&=&
\left|\frac{\tr_{\mathcal{H}}\left[e^{it\hat{H}_\tau^\textrm{eff}}e^{-(\beta+it)\hat{H}_0}\right]}{Z\left(\beta \right)}\right|^2,
\eeqa
This quantity  can  generally be extracted from the work distribution function associated with the general quench in (\ref{genquench}) via the inverse Fourier transform yielding the identity
\beqa
\mathcal{L}(t)
=\iint dWdW'p(W)p(W')\cos[(W-W')t].
\eeqa
Equivalently, the  work distribution function refers to the driving of the system from $\hat{H}_0$ to $\hat{H}_{\tau}$  when the dynamics is described by the  time-evolution operator  $\hat{U}(\tau)$.

\subsection*{Quantum work statistics of quantum chaotic systems}
\begin{figure}
\includegraphics[width = .95\columnwidth]{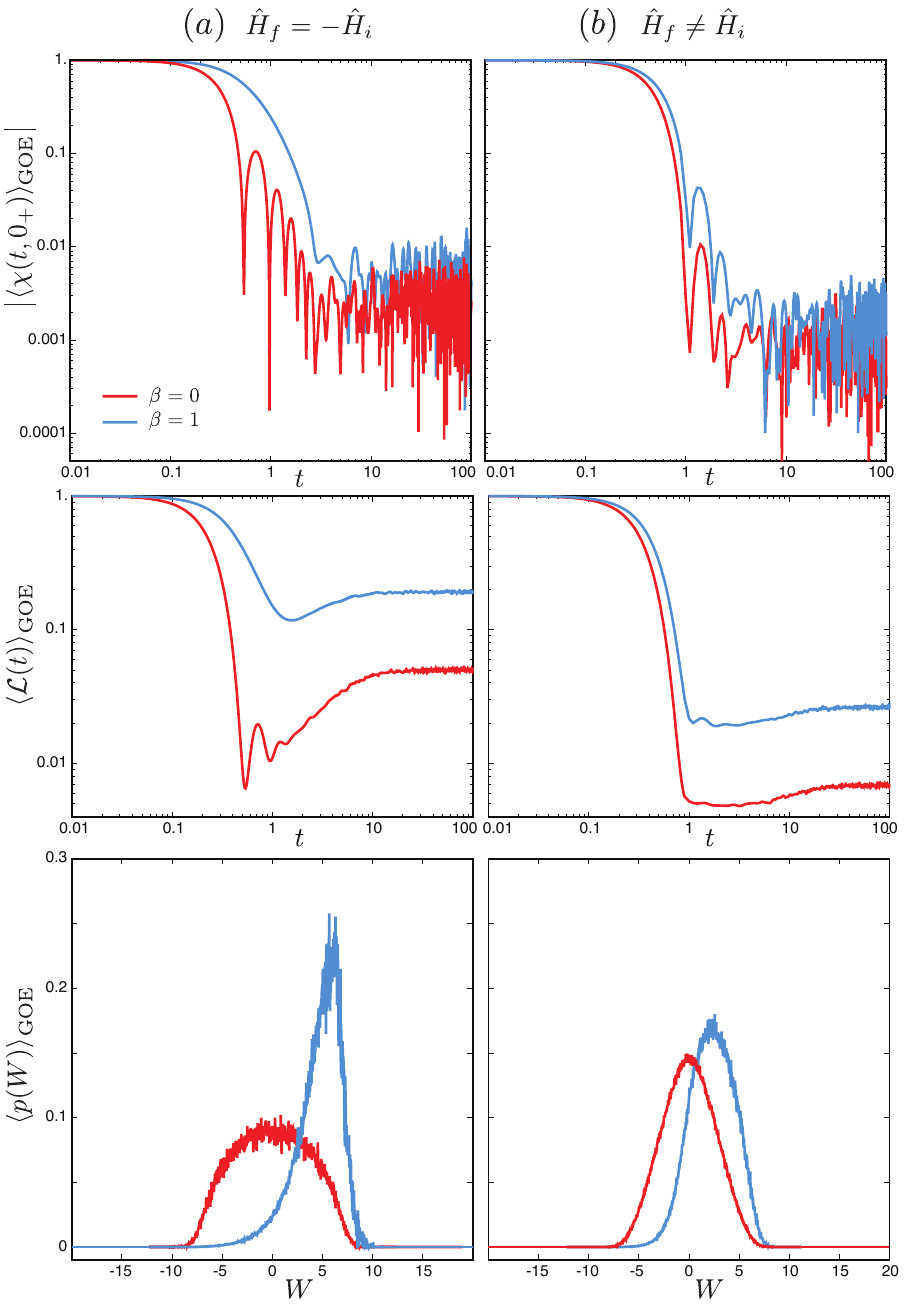}
\caption{{\bf Quantum work statistics of time reversal and information scrambling.} Absolute value of the characteristic function, $\left |\left\langle \chi(t,0^{+})\right\rangle _{\rm GOE}\right| $, generating the Loschmidt echo $\langle \mathcal{L}(t)\rangle_{\rm GOE} $ and work probability distribution $\langle p(W)\rangle_{\rm GOE}$ for a driven chaotic system implemented by sampling the initial and final Hamiltonians from the Gaussian  Orthogonal Ensemble, for (a) an exact time reversal of the Hamiltonian and (b) an arbitrary sudden quench. Results are averaged over 5000 realizations of the GOE for $N=20$ at infinite temperature ($\beta=0$, red) and $\beta=1$ (blue). \label{fig1}}
\end{figure}

We next  illustrate the relation between work statistics, Loschmidt echoes and information scrambling in a driven quantum chaotic system. The Hamiltonians we consider are random $N \times N$ Hermitian matrices sampled from  the Gaussian  Orthogonal Ensemble (GOE) \cite{MehtaBook}, which are invariant under  time reversal.
A sudden random quench can be implemented by choosing the initial and final Hamiltonians, $\hat{H}_0=\hat{H}_i$ and $\hat{H}_\tau=\hat{H}_f$, from  two independent GOE \cite{Lobejko2017a}.  
The corresponding characteristic function averaged over the GOE
\beqa \label{gfgue}
\left\langle \chi(t,0^{+})\right\rangle_{\rm GOE}&=&
\left\langle \frac{1}{Z(\beta)}{\rm Tr}\left( e^{-\sigma_f\hat{H}_f}\, e^{-\sigma_i\hat{H}_i}\right) \right\rangle_{{\rm GOE}}, 
\eeqa 
where $\sigma_i = \beta + it,\, \sigma_f = -it$,  is shown in Fig. \ref{fig1}.
The average work probability distribution, also presented in Fig. \ref{fig1}, then directly follows from Fourier transformation, Eq. (\ref{pWIF}), and reads  
\begin{equation} \label{eq:pWGOE}
\la p(W) \ra_{\rm GOE} =\frac{1}{2\pi}\int_{-\infty}^\infty dt \,    \left\langle \chi(t,0^+)\right\rangle_{\rm GOE} \: e^{-itW}. 
\end{equation}
In turn, the Loschmidt echo associated with this driven chaotic system follows from the equivalence identified above, Eq. (\ref{LEequivalence}),  
and can be evaluated from the characteristic function as 
\begin{equation} \label{LGOE}
\la \mathcal{L}(t)\ra_{\rm GOE} = \langle \, |\chi(t,0^+)|^2  \, \rangle_{{\rm GOE}}. 
\end{equation}

Figure \ref{fig1} shows the characteristic function,  Loschmidt echo and work probability distribution for (a) a sudden negation of the Hamiltonian $\hat{H}_f = -\hat{H}_i$, and (b) for an arbitrary sudden quench.  At infinite temperature, the work probability distribution is symmetric, representing the fact that the initial and final random Hamiltonians are drawn from ensembles with identical distributions. As the temperature is lowered, the initial thermal state samples predominantly the low-energy spectrum of the initial Hamiltonian, increasing the probability for trajectories associated with positive work, as manifested by the shift  of  $\la p(W)\ra_{\rm GOE}$  towards the positive real axis. Such a shift has a non-trivial effect on the decay of  $|\left\langle \chi(t,0^{+})\right\rangle _{\rm GOE}|^2$ whose long time behavior is characterized by higher values for increasing temperature.
Considering the identified equivalence with the survival amplitude, this decay can also be interpreted 
in terms of information scrambling, as explicit from the Loschmidt echo $\mathcal{L}_{\rm GOE}(t)$ represented in Fig. \ref{fig1}.
The Loschmidt echo associated with the work statistics for a perfect time reversal operation in one of the copies exhibits the following features, common to scrambling dynamics of many-body-chaotic systems \cite{Cotler2017b,Dyer2017a,delCampo2017a}: it reaches a minimum value at a dip,  followed by a ramp and a saturation at long values described by the long-time average. By contrast, the work statistics associated with a quench between two independent random Hamiltonians, leads to a Loschmidt echo characterized by an enhanced  decay as manifested by lower values of the dip, which is also broadened. In addition, the subsequent dynamics towards the long-time asymptotics no longer exhibits a clear ramp.  A comprehensive analysis of work statistics in quantum chaotic systems is presented in \cite{Chenu2018b}.

\section*{Conclusion}
In summary, we have shown a universal relation between the quantum work statistics and Loschmidt echo under arbitrary dynamics. Specifically, within the two projective energy measurement scheme,  we have shown that the generating function of the work probability distribution of an isolated quantum system prepared in a possible mixed state can be interpreted as the  Loschmidt echo amplitude of a purified density matrix in an enlarged Hilbert space, for a   quench acting on one of the copies. 
When the initial state is thermal, the Loschmidt echo describes the evolution of a thermofield double state and is ideally suited to assess information scrambling. In particular, the work statistics associated with the time-reversal operation -- the sudden negation of the system Hamiltonian -- is dictated by the analytic continuation of the partition function, recently proposed to diagnose quantum chaos. 
As a result, our work establishes a firm connection between the finite-time thermodynamics of closed quantum systems, irreversibility, and information  scrambling.





\section*{Acknowledgments}
It is a pleasure to thank John Goold, Juan Jaramillo, and Julian Sonner for insightful discussions.
Funding support from UMass Boston (project P20150000029279) and the John Templeton Foundation is  acknowledged. ILE acknowledges funding from Spanish MINECO/FEDER FIS2015-69983-P, the Basque Government through IT986-16, and UPV/EHU  with grant EHUA15/17.  JMV is supported by Ministerio de Econom\'{i}a y Competitividad FIS2015-69512-R and Programa de Excelencia de la Fundaci\'{o}n S\'{e}neca 19882/GERM/15. AC and AdC acknowledge the hospitality of the Simons Center for Geometry and Physics during  completion of this work.

\end{document}